\documentclass[%
 reprint,
superscriptaddress,
 amsmath,amssymb,
 aps,
prb,
]{revtex4-2}
\setcitestyle{numbers,square}
\usepackage{amsmath,amssymb}
\usepackage{tabularx}
\usepackage{graphicx}
\usepackage{bmpsize}
\usepackage{bm}
\usepackage{color}
\usepackage{amssymb}
\usepackage[version=3]{mhchem}
\usepackage{newtxmath}
\usepackage{physics}
\usepackage{hyperref}
\usepackage{mathtools}
\usepackage{xcolor}
\usepackage{ulem}
\usepackage{comment}

\newcommand{\red}[1]{\textcolor{red}{#1}}

\newcommand{\blue}[1]{\textcolor{blue}{ #1}}

\begin{document}
\title{Ferroelectricity in a magnon Bose-Einstein condensate: Nonreciprocal superfluidity, exceptional points, and Majorana bosons}
\author{Kazuki Yamamoto}
\affiliation{Department of Physics, The University of Osaka, Toyonaka, Osaka 560-0043, Japan}
\affiliation{Department of Physics, Kyoto University, Kyoto 606-8502, Japan}

\author{Takuto Kawakami}
\affiliation{Department of Physics, The University of Osaka, Toyonaka, Osaka 560-0043, Japan}
\affiliation{Center for Integrated Science and Humanities, Fukushima Medical University, Fukushima 960-1295, Japan}

\author{Mikito Koshino}
\affiliation{Department of Physics, The University of Osaka, Toyonaka, Osaka 560-0043, Japan}
\affiliation{Institute for Solid State Physics, The University of Tokyo, Kashiwa, Chiba 277-8581, Japan}

\begin{abstract}

We investigate a ferroelectric instability of a magnon Bose–Einstein condensate, mediated by its interaction with an electric field through a geometric Aharonov–Casher (AC) phase.
A distinct feature of the system is the positive feedback loop in which an electric field induces magnon orbital motion via the AC phase, generating electric polarization that in turn enhances the original field.
Based on bosonic Bogoliubov–de Gennes (BdG) mean-field theory, we show that this feedback drives a spontaneous ferroelectric transition in the magnon superfluid, accompanied by a persistent magnon supercurrent. 
In the resulting ferroelectric phase, the quasiparticle excitation spectrum becomes nonreciprocal, reflecting spontaneous breaking of spatial inversion symmetry. 
At the critical point of the transition, the bosonic BdG Hamiltonian exhibits a global coalescence of both eigenvalues and eigenvectors, forming exceptional points 
throughout the entire Brillouin zone.
The corresponding eigenstate is an equally weighted superposition of bosonic quasiparticle and quasihole states and is invariant under particle–hole transformation, allowing it to be interpreted as a bosonic analog of a Majorana fermion.

%
\end{abstract}
\maketitle


\section{Introduction}
Geometric phases are a fundamental concept in modern condensed matter physics.
A classic example is the Aharonov–Bohm (AB) phase, which arises in electronic systems subjected to a magnetic field.
This idea can be extended by considering the geometric phases for a magnetic dipole in an electric field, which is known as the Aharonov-Casher (AC) phase~\cite{Aharanov1984}. 

A representative physical system for studying the AC phase is a magnon system, where magnons are the quasiparticles associated with spin waves in magnetic insulators.
This system is particularly ideal because magnons possess a fixed magnetic dipole moment, and the AC phase is greatly enhanced by the strong spin–orbit coupling typically present in magnetic insulators~\cite{Katsura2005,Liu2011}.
In this context, an electric field acts as an effective vector potential for magnons~\cite{Meier2003,Nakata2015,SuWang2017,Nakata2017,Nakata2017_Landu_level,Owerre2018,Owerre_SciRep2018,Proskurin2019,Avishai2019,Li2020,Serha2023,Krivoruchko2024,WangYuanDongZhu2024,JinZhejunyu2024,Kuzmenko_PRB2025,Boliasova2025,WangYuanDong2025,Birnkammer2025,ChenQiHui2025,Yokoyama2025,LiDu2025}.
Consequently, several studies have proposed that such an effective vector potential can give rise to physical phenomena analogous to those observed for electrons in magnetic fields, including the Hall effect and the formation of Landau levels~\cite{Meier2003,Nakata2017_Landu_level}.

Recent investigations~\cite{Yamamoto2025} have uncovered fundamental differences between magnonic and electronic systems by analyzing their electromagnetic feedback mechanisms.
In electronic systems, the feedback is negative, as the orbital magnetization tends to weaken or counteract the external magnetic field.
Conversely, in magnonic systems, the feedback is positive: the electric polarization that is induced by a magnon orbital motion
strengthens the external electric field.

These fundamental differences become most apparent in the realm of superfluid physics.
In superconductors, complete diamagnetism, known as the Meissner effect, represents the extreme manifestation of the negative feedback in 
electronic systems.
By analogy, one may anticipate the emergence of phenomena corresponding to the extreme limit of positive feedback in
a magnon superfluid or a magnon Bose–Einstein condensate (BEC)~\cite{Demokritov2006,Nikuni2000,Ruegg2003,Giamarchi2008,Aczel2009,Sonin2010,Zapf2014,Nakata2014,Bozhko2016,Sun_PRL2016,Kimura2016,Kimura2017,Fjaerby_PRB2017,RuckriegelPRB2017,Yuan2018,Bozhko2019,Okuma_PRB_2019,Takei_PRB2019,Kimura2020,Sakurai2020,Olsson2020,Schneider_NatNanotech2020,Borisenko2020,Divinskiy2021,Kreil_PRB2021,Sukhachov_PRR2021,Nakata2021,Hayashida_PRR2021,Yun_PRB2023,Aftergood_PRL2023,Flavian_PRL2023,Lvov_PRB2024,Matsumoto_NatPhys2024,Esaki2024,Frostad_PRR2024,Nakata_PRR2024,Zhu2025,KhatuaPRB2025,FrostadPRB2025,ShengNatureMaterial2025,flynn_arXiv2025,Nomura_PRL2026}.
The magnon BEC has been experimentally observed in various settings, including field-induced magnon BEC in quantum magnets, for example, TlCuCl$_3$~\cite{Nikuni2000,Ruegg2003,Giamarchi2008,Aczel2009,Zapf2014,Kimura2016},
and pumped quasi-equilibrium magnon BEC, for example, yttrium iron garnet thin films~\cite{Demokritov2006,Bozhko2016,Sun_PRL2016,Bozhko2019,Divinskiy2021}.

In this work,
we investigate a ferroelectric instability in a magnon BEC coupled to electric fields via the AC phase. Using a bosonic Bogoliubov-de-Genne (BdG) mean-field theory~\cite{Shi1998,Biao2001,WuNiu2003,You_PRL2012,Zhang_PRA2013,ShindouPRB2013,McDonald2018,LieuPRB2018,LeinSatoPRB2018,KawabataPRX2019,Ohashi2020,Lyu_PRL2020,Julku_PRL2021,Julku_PRB2021,YokomizoMurakamiPRB2021,Wan_PRA2021,OkumaPRB2022,Jalali_PRL2023,Iskin_PRA2023,OkumaPRB2024,Guo_PRA2024,Tesfaye_PRR2025,Syljuaasen_PRB2025,Yu_npj2025,yuan2025}, we show that a positive electromagnetic feedback mechanism leads to spontaneous ferroelectric polarization in the absence of an external electric field, when the coupling to the electromagnetic field is strong enough. 
Specifically, an electric field induces magnon orbital motion via the AC phase, generating an electric polarization that in turn contributes to and enhances the electric field.
The resulting polarization breaks spatial inversion symmetry, reshapes the Bogoliubov excitation spectrum, and gives rise to non-reciprocal superfluidity.
At the critical point of the ferroelectric phase transition,
the Hamiltonian hosts exceptional points across the entire Brillouin zone, yielding a zero-energy state with completely flat dispersion. This eigenstate is an equal-weight superposition of bosonic quasiparticle and quasihole components and is invariant under particle–hole transformation. It can therefore be interpreted as a Majorana-like bosonic zero mode.
\begin{figure}[t]
    \centering
    \includegraphics{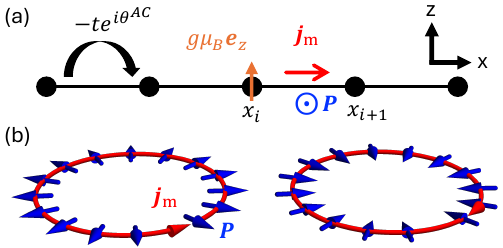}
    \caption{
    (a) Schematic of the one-dimensional magnon Bose–Hubbard model with AC phase.
    The chain lies along $x$-direction with lattice spacing $a=1$ and hopping $-t$.
    Magnons carry a magnetic dipole $\bm{\mu}=g\mu_B\vb{e}_z$ and are subject to an electric field $\vb{E}=E\vb{e}_y$, leading to an AC phase $\theta_\mathrm{AC}$ in nearest-neighbor hopping.
    The magnon supercurrent $\vb{j}_\mathrm{m}$ (red arrows along $x$-direction) induces an electric polarization $\vb{P}$ (blue arrows along $y$-direction), which generates a feedback electric field that further contributes to the AC phase.
    (b) Ferroelectric polarizations induced by persistent magnon supercurrent with $\Delta=\pm\Delta_0$.
    }
    \label{figure:schematic}
\end{figure}

\section{Model}

As a minimal theoretical description of a magnon BEC coupled to an electric field, we consider 
the one-dimensional magnon Bose–Hubbard model incorporating the AC phase [Fig.~\ref{figure:schematic}(a)].
The Hamiltonian for this model is expressed as
\begin{align}
    H_0=-t\sum_{i}&(e^{-i\theta_\mathrm{AC}}b_i^\dagger b_{i+1}+e^{+i\theta_\mathrm{AC}}b_{i+1}^\dagger b_i)
    +\frac{U}{2}\sum_in_i^2,\label{eq: Bose-Hubbard model}
\end{align} 
where site indices $i$ range from 1 to $N$, and we impose periodic boundary condition.
$b_i$ annihilates a magnon and
$n_i\equiv b_i^\dagger b_i$ counts the number of magnons in position $x_i=ia$, where $a$ denotes a lattice constant.
$t>0$ denotes a nearest-neighbor hopping amplitude,
and $U>0$ represents the onsite repulsive interaction.
This model can be derived as an effective theory from a class of microscopic spin models.
As an illustrative example, we demonstrate the derivation of Eq.~\eqref{eq: Bose-Hubbard model} in Appendix.~\ref{app: derivation of spin Hamiltonian} for a spin-$1/2$ antiferromagnetic quantum dimer magnet, a minimal model of TlCuCl$_3$-type materials.
Although we assume $t>0$, 
our conclusions, including the emergence of ferroelectricity, are insensitive to a sign of $t$.


Magnons have a fixed magnetic dipole moment $\vb*{\mu}=g\mu_B\vb{e}_z$,
where $g\simeq 2$ is the Lande g-factor and $\mu_B>0$ is the Bohr magneton.
Thus, they interact with the electric field $\vb{E}=E\vb{e}_y$ through the AC phase~\cite{Aharanov1984} 
\begin{equation}
    \theta_\mathrm{AC}=-\frac{1}{\hbar}\int_i^{i+1}\vb{A}_{\mathrm{m}}\cdot d\vb{r},\label{eq: definition of AC phase}
\end{equation}
where 
\begin{equation}
    \vb{A}_\mathrm{m}=g_\mathrm{AC}^{}\vb{E}\times\vb{e}_z\label{eq: definition of Am}
\end{equation}
is the effective vector potential, pointing along the $x$-direction.
Here, the constant parameter $g_\mathrm{AC}~{}$ represents the strength of a spin-orbit coupling.
In vacuum, it amounts to $g\mu_B/c^2$, however, it greatly enhances
in typical magnetic insulators~\cite{Katsura2005,Liu2011}.
A simplified derivation of AC phase can also be found in Ref.~\cite{Yamamoto2025}.

Assuming the approximate U(1) spin conservation,
the magnon current density $\vb{j}_\mathrm{m}=j_{i,i+1}/a^2~\vb{e}_x$ is defined to satisfy the continuity equation
$\partial_t n_i=i/\hbar[H_0,n_i]\equiv j_{i-1,i}-j_{i,i+1}$, giving
\begin{equation}
    j_{i,i+1}=\frac{it}{\hbar}(e^{i\theta_\mathrm{AC}}b_{i+1}^\dagger b_i-e^{-i\theta_\mathrm{AC}}b_{i}^\dagger b_{i+1}).\label{eq:current operator}
\end{equation}
This definition agrees with the expression 
obtained from the functional derivative of the Hamiltonian with respect to the effective vector potential.
Physically, this magnon current is responsible for generating the electric dipole $\vb{P}_{i,i+1}=-\partial H_0/\partial \vb{E}$.
The electric polarization $\vb{P}=\vb{P}_{i,i+1}/a^3$, i.e., electric dipole per volume, is related to the magnon current density through the general relation ~\cite{Yamamoto2025,Katsura2005,Liu2011,Proskurin_PRB2018,Fujiwara_PRB2023,Go2024,To2025,Tang2025},
\begin{equation}
    \vb{P}
    =g_\mathrm{AC}~{}\vb{j}_\mathrm{m}\times\vb{e}_z.\label{eq:P}
\end{equation}
The proportionality~[Eq.~\eqref{eq:P}] between electric polarization and magnetic dipole current is complementary to the relation~[Eq.~\eqref{eq: definition of Am}] between effective vector potential and electric field~\cite{Yamamoto2025}.
Through the basic relation $\epsilon_0\vb{E}=\vb{D}-\vb{P}$ in electrodynamics,
the electric polarization contributes to the total electric field,
which in turn produces an additional self-induced AC phase.
The key feature of our model [Eq.~\eqref{eq: Bose-Hubbard model}] is that the Peierls AC phase $\theta_\mathrm{AC}$ is dynamically determined through feedback from the state of the system.
While in a different context, a related idea has been explored in Ref.~\cite{Takasan2024}, where the hopping amplitude in the bosonic lattice model depends on the state of the system.

Building on these considerations, we seek the total energy of the system, explicitly incorporating the energy of the electromagnetic field. 
In accordance with the standard thermodynamic treatment of electromagnetic fields,
the total Hamiltonian $H[\vb{D}]$ under an external electric field $\vb{D}$ is obtained as
\begin{align} 
    H[\vb{D}] 
    &=\int_{0}^{\vb{D}}\int d\vb{r}\,\, \vb{E}\cdot \delta \vb{D}
    \notag\\
    &=\int_{0}^{\vb{D}}\int d\vb{r}\, 
    \Biggl(
     - \vb{P}\cdot \delta\vb{E}
    +\frac{\vb{D}\cdot \delta\vb{D}-\vb{P}\cdot \delta \vb{P}}{\varepsilon_0} 
     \Biggr)
     \notag\\
    &= H_0+\int d\vb{r}\,\,\frac{\vb{D}^2-\vb{P}^2}{2\varepsilon_0},\label{eq: H[D]}
\end{align}
where we have used the equation $\epsilon_0\vb{E}=\vb{D}-\vb{P}$ in the second equality and $\vb{P}_{i,i+1}=-\partial H_0/\partial \vb{E}$ in the third.
The same approach was used in Ref.~\cite{Yamamoto2025} to derive the total free energy of general magnetic-dipole superfluids in the phenomenological Ginzburg-Landau theory.


For the present system,
we set $\vb{D}=0$ in Eq.~\eqref{eq: H[D]}
to examine the intrinsic behavior of the system without the influence of an external electric field. 
The total Hamiltonian is then expressed as
\begin{equation}
    H=H_0-\frac{g_\mathrm{AC}^{2}}{2\epsilon_0a}\sum_ij_{i,i+1}^2,\label{eq:total Hamiltonian}
\end{equation}
where we have used $P\propto j_\mathrm{m}= j_{i,i+1}/a^2$~[Eq.~\eqref{eq:P}].

\begin{figure}[t]
    \centering
    \includegraphics{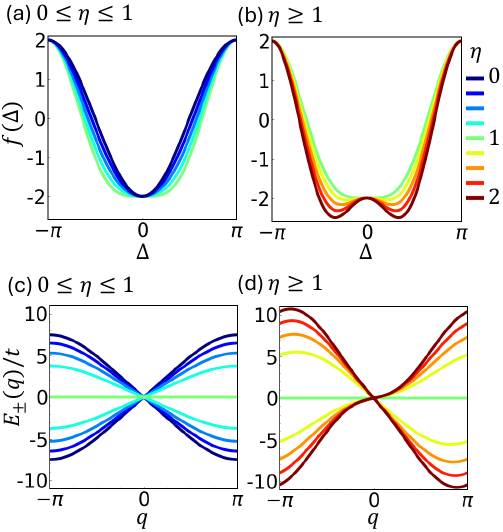}
    \caption{
    (a),(b): The normalized single-particle mean-field energy $f(\Delta)$ is shown as a function of $\Delta$, with color representing different values of $\eta$.
    (c),(d): The Bogoliubov spectrum of the bosonic quasiparticle $E_+(q)=E(q)$ and quasihole $E_-(q)=-E(-q)$ are presented, where we set $u=5$.
    A Majorana boson emerges exactly at the critical point $\eta=1$.
    }
    \label{figure:ground state energy}
\end{figure}

\section{Ferroelectricity in magnon BEC}
Owing to the discrete translational symmetry of the system, the Bloch wave vector 
serves as a good quantum number.
It is then reasonable to assume that condensation occurs in a lowest-energy single-particle Bloch state~\cite{Shi1998},
\begin{equation}
    \expval{b_i}=\sqrt{n_0}e^{ik_cx_i},\label{eq:Bloch ansatz}
\end{equation}
where $n_0$ represents the number of condensed bosons per site.
The optimal value of $-\pi/a<k_c<\pi/a$ is chosen to minimize the mean-field energy $\expval{H}$, as described below.

Substituting the ansatz [Eq.~\eqref{eq:Bloch ansatz}] into the current operator [Eq.~\eqref{eq:current operator}] yields the magnon supercurrent
$\expval{j_{i,i+1}}=(2tn_0/\hbar)\sin\Delta$,
where
\begin{equation}
    \Delta\equiv k_ca-\theta_\mathrm{AC}.\label{eq:definition of Delta}
\end{equation}
The magnon supercurrent $\expval{\vb{j}_\mathrm{m}}$ produces the electric polarization by Eq.~\eqref{eq:P}, which is an additional electric field.
Since we do not apply the external electric field, the AC phase is determined solely from this contribution, giving $\theta_\mathrm{AC}=-\eta\sin\Delta$,
where 
\begin{equation}
    \eta=\frac{g_\mathrm{AC}^{2}}{m^*\varepsilon_0a^3}n_0
    \label{eq:definition_eta}
\end{equation}
is the dimensionless parameter~\cite{Yamamoto2025} that characterizes the strength of the spin-orbit coupling,
and $m^*=\hbar^2/(2ta^2)$ is an effective mass for a non-interacting magnon.
Thus, Eq.~\eqref{eq:definition of Delta} can be recast as
\begin{equation}
    k_ca=\Delta-\eta\sin\Delta,\label{eq:kc_and_Delta}
\end{equation}
thereby fixing the value of $k_c$ for a given $\Delta$.

To identify the lowest-energy single-particle Bloch state, we evaluate the mean-field energy by substituting the ansatz [Eq.~\eqref{eq:Bloch ansatz}] into the total Hamiltonian [Eq.~\eqref{eq:total Hamiltonian}], giving
\begin{equation}
    \expval{H}
    =-2tn_0\cos\Delta+\frac{U}{2}n_0^2-\eta tn_0\sin^2\Delta
    \label{eq: ground state energy}.
\end{equation}
The mean-field energy $\expval{H}$, given in Eq.~\eqref{eq: ground state energy}, depends on $k_c$ only through $\Delta$. 
Thus, we begin by minimizing it with respect to $\Delta$, after which we determine the condensate wave vector $k_c$.
Eq.~\eqref{eq: ground state energy} can be 
simplified to a dimensionless form
\begin{equation}
    f(\Delta)=-2\cos\Delta-\eta\sin^2\Delta
\end{equation}
where a constant term has been neglected. 

In Fig.~\ref{figure:ground state energy}~(a),(b), we plot $f(\Delta)$ for different regimes of $\eta$. When $0 \leq \eta \leq 1$, the ground-state energy has a minimum at $\Delta = 0$, where both the magnon supercurrent $j_\mathrm{m} (\propto\sin\Delta)$ and the associated electric polarization $P (\propto j_\mathrm{m})$ vanish. 
By contrast, for $\eta \geq 1$ the ground-state energy develops minima at finite values $\Delta = \pm \Delta_0$, with $\Delta_0 \equiv \arccos(1/\eta)$, giving rise to a persistent magnon supercurrent and an emergent electric polarization proportional to $\sin\Delta \neq 0$. 
Correspondingly, the condensate Bloch wave vector $k_c$ in Eq.~\eqref{eq:kc_and_Delta} acquires a finite, nonzero value.
This behavior represents ferroelectricity accompanied by spontaneous breaking of spatial inversion symmetry.
The two solutions with $\Delta = \pm \Delta_0$ are related by the parity operation and correspond to states with opposite directions of electric polarization, as illustrated in Fig.~\ref{figure:schematic}(b). In the following, we adopt $\Delta = +\Delta_0$ as the ground-state configuration.
Although we assume $t>0$, the results also hold for $t<0$, which is accounted for by the substitution $f(\Delta)\rightarrow f(\Delta+\pi)$.

The one-dimensional model we have studied should be interpreted as an effective description of quasi-one-dimensional or anisotropic three-dimensional systems, such as weakly coupled spin chains in quantum magnets. In realistic materials, weak interchain couplings play an essential role in stabilizing long-range order, which justifies the use of a mean-field approach. Accordingly, our analysis implicitly assumes a three-dimensional array of weakly coupled chains, and the central results, particularly the feedback-driven ferroelectric phase transition, are expected to remain robust in such realistic settings.

In Appendix.~\ref{sec: Bohr-van Leuwen}, we develop a magnonic analog of Bohr-van Leeuwen theorem: classical magnonic systems in thermal equilibrium are incapable of producing a net electric polarization.
This demonstrates that the ferroelectric polarization in magnon BEC presented in this work is attributed to intrinsically quantum mechanical origins.




\section{Bosonic BdG Hamiltonian}
We apply the Bogoliubov approximation~\cite{Shi1998} to examine the stability and excitation properties of the magnon BEC.
The bosonic field operator is decomposed in the Bloch basis as
$b_i=\expval{b_i}+\delta b_i$, where
\begin{align}
    \delta b_i=\frac{1}{\sqrt{N}}\sum_{q\ne0}e^{i(k_c+q)x_i}
    \delta b_{k_c+q},\label{eq:expand in fluctuation}
\end{align}
which corresponds to the fluctuation from the ground-state configuration.
By substituting Eq.~\eqref{eq:expand in fluctuation} into the total Hamiltonian $H - \mu N$ and retaining terms up to quadratic order in $\delta b_i$, we obtain the bosonic BdG Hamiltonian
%
$H_\mathrm{BdG}=1/2\sum_{q\ne0}\psi_q^\dagger H_\mathrm{B}(q)\psi_q$,
where $\psi_q=(\delta b_{k_c+q},\delta b^\dagger_{k_c-q})^t$ denotes the Nambu spinor and
\begin{equation}
    H_\mathrm{B}(q)=
    \mqty(h(q)&s(q)\\s^\ast(-q)&h^\ast(-q)),\label{eq:bosonic BdG Hamiltonian}
\end{equation}
where
\begin{align}
    h(q)&=-2t\cos(\Delta+q)+2Un_0-\mu\notag\\
    &\quad\quad-\eta t(1+\cos q-2\cos(2\Delta+q)),\notag\\
    s(q)&=Un_0-\eta t(\cos q-\cos 2\Delta).
\end{align}

The excitation spectrum is given by the poles of the Green's function $G(q,i\omega)=[i\omega\sigma_z-H_\mathrm{B}(q)]^{-1}$. 
Since the excitation spectrum must be gapless at $q = 0$ \cite{Shi1998}, we impose the condition $\det H_\mathrm{B}(q=0) = 0$, which determines the chemical potential $\mu$ as
\begin{equation}
    \mu=
    \begin{cases}
    -2t+Un_0  & \text{if $0\le\eta\le1$,} \\
    -2t\cos\Delta+Un_0-2t\eta\sin^2\Delta  & \text{if $\eta\ge1$.}
    \end{cases}\label{eq:mu}
\end{equation}
Alternatively, the same equation can be obtained from the thermodynamic relation $\mu = \partial \langle H \rangle / \partial n_0$, with $\langle H \rangle$ defined in Eq.~\eqref{eq: ground state energy}, by noting that $\eta$ depends on $n_0$ as given in Eq.~\eqref{eq:definition_eta}.

\section{Bogoliubov excitation spectrum}
We calculate the Bogoliubov excitation spectrum by diagonalizing the pseudo-Hermitian matrix~\cite{Ashida2020,Deng_PRL2025}
\begin{equation}
    L(q)\equiv\sigma_z H_\mathrm{B}(q)
    =\sigma_zL^\dagger(q)\sigma_z,\label{eq:L(q)}
\end{equation}
that is generally a non-Hermitian matrix, despite the fact that 
$H_\mathrm{B}(q)$ itself is a Hermitian matrix.
The eigenvalue equations are expressed as
\begin{align}
    L(q)\ket{\psi_q^\pm}&=\pm E(\pm q)\ket{\psi_q^\pm}
\end{align}
where the eigenstates $\ket{\psi_q^\pm}$ represent the quasiparticle and quasihole states, respectively.
They are related by $\ket{\psi_{q}^-}=\sigma_x\ket{\psi_{-q}^+}^*$ because of the particle-hole symmetry
\begin{equation}
    \sigma_xL^*(-q)\sigma_x=-L(q).\label{eq:PHS}
\end{equation}


In Fig.~\eqref{figure:ground state energy}~(c),(d), the Bogoliubov band structures are presented, where we set $u\equiv Un_0/t=5$.
The different colors represent various values of $\eta$.

When $0<\eta<1$ (blue bands in Fig.~\ref{figure:ground state energy}(c)), 
we reproduce the typical behaviors of the conventional superfluids well known in the literature~\cite{Shi1998},
which has the gapless linear dispersion around $q=0$.
Up to this point, the system respects a space-inversion symmetry and satisfies $E(q)=E(-q)$.
As $\eta$ increases from 0 to 1, both the quasiparticle band and the quasihole band gradually move toward zero.

When $\eta=1$ (a green band in Fig.~\ref{figure:ground state energy}(c),(d)), the quasiparticle and quasihole bands merge into a fully degenerate zero-energy state with a completely flat dispersion.
This is precisely the critical point where the ferroelectric phase transition occurs.
Importantly, this zero-energy state is not a simple degeneracy of eigenvalues, but an exceptional point where the eigenvectors also coalesce. 
The pseudo-Hermitian matrix [Eq.~\eqref{eq:L(q)}] takes the form
\begin{equation}
    L(q)\propto
    \mqty
    (1&1\\
    -1&-1)\label{eq: L(q) when eta=1}
\end{equation}
throughout the entire Brillouin zone, with a proportionality factor $Un_0+t(1-\cos q)\ne0$.
For each $q\ne0$, Eq.~\eqref{eq: L(q) when eta=1} possesses only a single eigenvector $(1,-1)^t$ and is therefore not diagonalizable, which is the defining feature of an exceptional point.
Notably, such exceptional points are not isolated in momentum space but extend over the entire Brillouin zone.

The resulting zero-mode is an equally weighted superposition of quasiparticle and quasihole components and is invariant under particle–hole transformations. In this sense, its structure is directly analogous, at the level of symmetry, to a Majorana fermion. This motivates us to refer to it as a ``Majorana boson''.
At the same time, this mode is a zero mode at the critical point and should be understood as a dynamically unstable mode associated with enhanced fluctuations, rather than a stable quasiparticle like the Majorana fermions in topological superconductors~\cite{leijnse2012,alicea2012,sato2016,mizushima2016,YamamotoPRB2026}. Despite this, the mode is of particular interest because its self-conjugate structure originates from the exceptional-point behavior of the non-Hermitian bosonic BdG Hamiltonian $L(q)$. This mechanism is distinct from that of conventional dynamical instabilities and may lead to qualitatively different critical behavior~\cite{Hanai_PRR2020}.
We note, in previous works, Majorana bosons were identified in dissipative, non-Hermitian settings~\cite{Flynn2021,Flynn_PRB2023}, whereas here Majorana-like bosonic zero modes emerge at a critical point of a phase transition in a Hermitian setting due to an effective non-Hermitian nature of the bosonic BdG Hamiltonian.



When $\eta>1$ (red bands in Fig.~\ref{figure:ground state energy}(d)), the quasiparticle band and the quasihole band start to separate again.
However, this separation is accompanied by a notable change: the Bogoliubov spectrum becomes non-reciprocal $E(q)\ne E(-q)$.
This behavior is naturally understood as a manifestation of a space-inversion symmetry breaking, which originates from the development of spontaneous electric polarization.
\begin{figure}[t]
    \centering
    \includegraphics{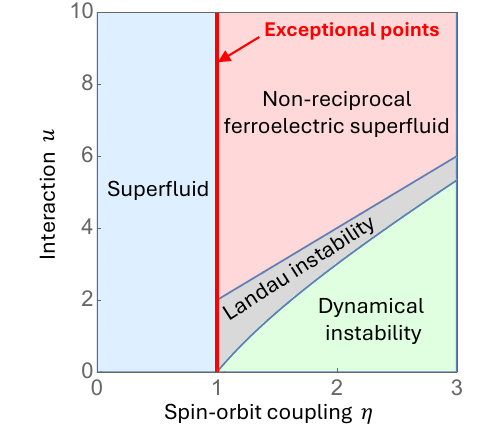}
    \caption{
    Phase diagram for the magnon BEC. 
    Ferroelectric phase transition occurs at $\eta=1$, corresponding to exceptional points.
    The Landau instability occurs at $u=2\eta$ and the dynamical instability occurs at $u=2(\eta-1/\eta)$.
    }
    \label{figure:Phase_diagram}
\end{figure}

\section{Landau and dynamical instabilities}
We have shown that, when $\eta>1$, the spontaneous magnon supercurrent is induced and the ferroelectricity appears. 
However, for the ground state to remain stable, the condition $E(q)\ge0$ must hold for every $q$.
Depending on the relationship between the parameters $u$ and $\eta$, the system might exhibit instabilities.
This can be seen by focusing on the low-energy ($q\approx0$) behavior of the energy eigenvalue spectrum
\begin{equation}
    E(q)\approx
    \begin{cases}
    t\sqrt{2u(1-\eta)}|q|  & \text{if $0\le\eta\le1$,} \\
    2t\sqrt{\eta-\frac{1}{\eta}}\qty(-\sqrt{\frac{1}{\eta}}q+\sqrt{\frac{u-2(\eta-\frac{1}{\eta})}{2}}|q|)  & \text{if $\eta\ge1$.}
    \end{cases}
\end{equation}

The ground state is stable for all $u>0$ when $0\le\eta\le1$.
In contrast, the ground state may become unstable when $\eta\ge1$.
More precisely, as $u$ decreases, $E(q)$ first becomes negative and eventually acquires a complex value.

When $u<2\eta$, the energy eigenvalue becomes negative, leading to Landau instability~\cite{Biao2001,WuNiu2003}. 
In this regime, creating a quasiparticle lowers the total energy of the system, so the system spontaneously emits excitations. 
As a result, the superfluid can no longer sustain its flow, and the flow becomes dissipative.
It should be noted that the Landau instability in our system exhibits a qualitatively different character from the conventional case. In usual superfluids, the Landau instability refers to the breakdown of a metastable current-carrying state, while the true ground state remains non-current-carrying. In contrast, in our system the electromagnetic feedback drives the formation of a spontaneous current-carrying ground state, in which the condensate acquires a finite momentum without any external driving. Importantly, this self-organized superflow can intrinsically exceed the conventional critical velocity, leading to a Landau instability of what would otherwise be the ground state. In this sense, the instability is not induced by externally imposed flow, but instead arises as an inherent consequence of the self-consistent coupling between magnon current and electric polarization.

When $u<2(\eta-1/\eta)$, the energy eigenvalues become complex, signaling dynamical instability~\cite{Biao2001,WuNiu2003}.
This is a stronger instability condition than the Landau instability. 
The imaginary part of the eigenvalue plays the role of a growth or decay rate, resulting  in the collapse or fragmentation of the condensate.
This condition can also be derived in an alternative manner using the ground-state energy in Eq.~\eqref{eq: ground state energy}. 
Since the third term in Eq.~\eqref{eq: ground state energy} is proportional to $n_0^2$ (noting that $\eta \propto n_0$), the expression can be rewritten as
$\expval{H}=-2tn_0\cos\Delta+U_\mathrm{eff}n_0^2/2$,
where
\begin{equation}
    U_\mathrm{eff}\equiv U-\frac{2t\eta}{n_0}\sin^2\Delta\label{eq:definition_of_U_eff}
\end{equation}
represents an effective interaction between bosons.
The condition for the dynamical instability coincides with the region where the effective interaction becomes attractive, i.e., $U_\mathrm{eff}<0$.
Thus, the electromagnetic feedback can qualitatively modify the effective interaction between magnons. In particular, even when the bare magnon–magnon interaction is repulsive, the feedback mechanism can render the effective interaction attractive in the current-carrying state. As a result, a dynamical instability can emerge in a regime where, in conventional lattice boson systems with purely repulsive interactions, the superflow would remain stable. This feedback-induced sign reversal of the effective interaction is absent in standard superflow problems and represents a fundamentally new mechanism.

Under these instabilities, the mean-field ansatz based on a Bloch wave function [Eq.~\eqref{eq:Bloch ansatz}] is no longer valid.
A detailed investigation of this regime lies beyond the scope of the present work.
Importantly, in the case of magnon BEC in quantum dimer magnets (see Appendix.~\ref{app: derivation of spin Hamiltonian}), the magnons are described as hard-core bosons, which correspond to an effectively infinite on-site repulsive interaction $U\rightarrow\infty$.
This strong repulsion ensures that the system remains stable against the aforementioned instability.
Fig.~\ref{figure:Phase_diagram} presents a phase diagram that summarizes the results of this study.

\section{Conclusion}
Within a minimal model, we have investigated a feedback-driven ferroelectric quantum phase transition in a magnon Bose–Einstein condensate, arising from its coupling to electric fields via the geometric Aharonov–Casher phase. A key feature of this system is a positive feedback loop: an applied electric field induces magnon orbital motion through the AC phase, which generates an electric polarization that further reinforces the original field.
Using a mean-field approach, we constructed the phase diagram as a function of the repulsive interaction strength $u$ and the spin–orbit coupling $\eta$. In particular, for $\eta>1$, the system exhibits ferroelectricity as a direct consequence of this feedback mechanism. We also find that the breaking of spatial inversion symmetry leads to nonreciprocal Bogoliubov quasiparticles.
At the critical point, the transition is characterized by global exceptional points and the emergence of a Majorana-like bosonic excitation, reflecting the intrinsically non-Hermitian nature of the underlying bosonic Bogoliubov–de Gennes Hamiltonian.


We note that ferroelectricity has been experimentally observed for the magnon BEC in the quantum dimer magnet TlCuCl$_3$~\cite{Kimura2016}.
Although a quantitative comparison with experiment remains for future study,
we estimate the dimensionless parameter $\eta$ [Eq.~\eqref{eq:definition_eta}] for the magnon BEC in TlCuCl$_3$ in order to assess the feasibility of the proposed mechanism.
Using representative material parameters from the literature, we take the lattice constant as $a=3.98$~\AA~\cite{YamadaJPSJ2008}, 
the effective magnon mass as $m^*=2.61\times 10^{-29}~\mathrm{kg}$~\cite{YamadaJPSJ2008},
and the number of condensed bosons per site as $n_0=0.5$, which is a reasonable value for hardcore bosons with a maximum occupancy of unity.
The strength of the AC effect in typical magnetic insulators is characterized by the parameter $g_\mathrm{AC}\approx g\mu_B/c^2\times 10^6$~\cite{Katsura2005,Liu2011}, implying that the coupling is enhanced by six orders of magnitude compared to its value in vacuum.
Substituting these values into Eq.~\eqref{eq:definition_eta}, we obtain $\eta\approx1.5$.
This result indicates that $\eta$ exceeds unity, suggesting that the system lies within a favorable regime where the proposed ferroelectric instability is expected to be significant.
\\\indent
In addition, the model we propose is intentionally minimal and general, and in principle applies not only to magnon superfluids but also to a broader class of magnetic dipole superfluids. In this context, artificial platforms like cold-atom systems in optical lattices could provide a potential platform to observe the predicted effects, provided that $g_\mathrm{AC}^{}$ can be engineered and tuned appropriately~\cite{Slim2024,Busnaina2024}.

\section*{Acknowledgment}
We acknowledge fruitful discussions with Takeshi Mizushima and Ryo Hanai.
This work was supported in part by JSPS KAKENHI Grants No. JP20H01840, No. JP20K14415, No. JP21H05236, No. JP21H05232, No. 23KJ1518, No. 24K06921and by JST CREST Grant No. JPMJCR20T3, Japan.
\appendix

\begin{figure}[t]
    \centering
    \includegraphics{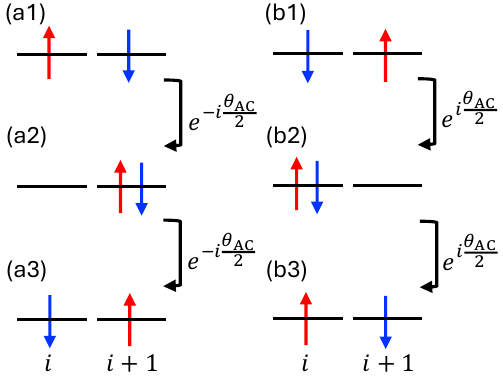}
    \caption{
    Schematic illustration of the superexchange processes in the spin-orbit coupled Hubbard model, assuming half-filling and strong Coulomb repulsion.
    (a1)–(a3) Virtual hopping processes in $S_i^-S_{i+1}^+$ term.
    An up-spin electron initially at site $i$ hops to site $i+1$, creating a doubly occupied intermediate state with energy cost $U_\mathrm{el}$. 
    Subsequently, a down-spin electron hops from site $i+1$ to site $i$, restoring single occupancy.
    As a result of this sequence, a spin $S^z=-1$ (i.e. magnetic dipole $g\mu_B\vb{e}_z$) is effectively transferred from site $i+1$ to site $i$.
    (b1)–(b3) Complementary virtual processes in $S_i^+S_{i+1}^-$ term.
    This process leads to an effective transfer of spin $S^z=-1$ 
    from site $i$ to site $i+1$.
    The two classes of processes, (a) and (b), acquire opposite AC phases due to their opposite directions of spin transport. 
    }
    \label{figure: superexchange}
\end{figure}


\section{Microscopic derivation of magnon Bose-Hubbard model with AC phase}\label{app: derivation of spin Hamiltonian}

In the main text, we introduced a magnon Bose–Hubbard model with an Aharonov–Casher (AC) phase [Eq.~\eqref{eq: Bose-Hubbard model}] as a minimal framework to investigate the dielectric properties of a magnon Bose–Einstein condensate (BEC). This model arises as an effective description of a broad class of microscopic spin systems. As a representative example, in this Appendix we present its derivation for a spin-$1/2$ antiferromagnetic quantum dimer magnet, which serves as a minimal model for TlCuCl$_3$-type systems~\cite{Nikuni2000,Ruegg2003,Giamarchi2008,Aczel2009,Zapf2014,Kimura2016}.


In Appendix~\ref{subapp: AC phase}, we start from  an electronic Hubbard model with spin–orbit coupling in the presence of an electric field.
By a standard perturbative treatment, we obtain an antiferromagnetic spin-1/2 Heisenberg model in which the electric field enters through the AC phase~\cite{Katsura2005,Liu2011}.


In Appendix~\ref{subapp: spin model}, following previous works~\cite{Nikuni2000,Ruegg2003,Giamarchi2008,Aczel2009,Zapf2014,Kimura2016}, we introduce a concrete minimal spin model and derive the one-dimensional magnon Bose–Hubbard model [Eq.~\eqref{eq: Bose-Hubbard model}]. Applying the argument developed in Appendix~\ref{subapp: AC phase}, we show that magnon hopping processes acquire the Aharonov–Casher (AC) phase in the presence of an electric field.
We also discuss how these magnons undergo Bose–Einstein condensation under an applied magnetic field.

\subsection{Spin-1/2 antiferromagnetic Heisenberg model with AC phase}\label{subapp: AC phase}
We consider the electronic Hubbard model with spin–orbit coupling in the presence of an external electric field.
Assuming half-filling and strong Coulomb repulsion, we derive an effective spin-$1/2$ antiferromagnetic Heisenberg model that interacts with electric field via the Aharonov–Casher (AC) phase.

We consider a one-dimensional lattice aligned along the $x$-direction, with the electric field applied along the $y$-direction. 
The Hamiltonian is then given by
\begin{align}
    H_\mathrm{el}&=-t_\mathrm{el}\sum_{i}\sum_{\sigma=\uparrow,\downarrow}
    \Bigl[~
    e^{ie/\hbar\int_i^{i+1}\vb{A}\cdot d\vb{r}}
    c^\dagger_{i,\sigma}c_{i+1,\sigma}\notag\\
    &+
    e^{-ie/\hbar\int_i^{i+1}\vb{A}\cdot d\vb{r}}
    c^\dagger_{i+1,\sigma}c_{i,\sigma}
    \Bigr]
    +U_\mathrm{el}\sum_i n_{i,\uparrow}n_{i,\downarrow}
\end{align}
where $t_\mathrm{el}>0$ is the nearest-neighbor hopping amplitude,
$U_\mathrm{el}>0$ is the on-site Coulomb repulsion.
The operator $c_{i\sigma}^\dagger\,( c_{i\sigma})$ creates (annihilates) an electron with electric charge $-e$ and spin $\sigma=\uparrow,\downarrow$ at lattice site $i$, and
$n_{i\sigma}=c_{i\sigma}^\dagger c_{i\sigma}$.

The effect of the spin-orbit coupling is introduced as a Peierls phase of a non-Abelian SU(2) vector potential~\cite{Avishai2019},
\begin{align}
    e\vb{A}&=\frac{1}{c^2}\vb{E}\times\bm{\mu}_e\notag\\
    &=-\frac{1}{2}g_\mathrm{AC}^{}\vb{E}\times\bm{\sigma},\label{eq: SU(2) vector potential}
\end{align}
where $\bm{\mu}_e=-g\mu_B\bm{\sigma}/2$ is the magnetic-dipole moment of the electron spin, and $g_\mathrm{AC}^{}=g\mu_B/c^2$ in vacuum.
This vector potential~[Eq.~\eqref{eq: SU(2) vector potential}] represents the spin-orbit coupling. 
To see this, let $\vb{k}$ denote the electron momentum.
We then obtain the relation $\vb{k}\cdot\vb{A}\propto\bm{\mu}_e\cdot(\vb{k}\times\vb{E})$.
This expression takes the form of a Zeeman coupling, indicating that $\vb{B}_\mathrm{eff}\propto-(\vb{k}\times\vb{E})$ can be interpreted as an effective magnetic field experienced by the electron in its rest frame~\cite{Avishai2019}.
The $g_\mathrm{AC}^{}$ is known to be enhanced by $\sim10^6$ in typical magnetic insulators~\cite{Katsura2005,Liu2011}.

When an electron hops from site $i$ to $i+1$ along the $x$-direction, its spin state acquires a Peierls phase of the SU(2) vector potential [Eq.~\eqref{eq: SU(2) vector potential}].
In the spin basis $\sigma=\uparrow,\downarrow$, which consists of eigenstates of the spin $z$-component,
the hopping direction ($x$-direction) is orthogonal to both the electric field ($y$-direction) and the magnetic dipole moment ($z$-direction).
Thus, the resulting Peierls phase can be expressed as
\begin{align}
    \exp[-i\frac{e}{\hbar}\int_i^{i+1}\vb{A}\cdot d\vb{r}]
    =\mqty
    (e^{-i\theta_\mathrm{AC}/2}&0\\
    0&e^{i\theta_\mathrm{AC}/2}),
\end{align}
which is a diagonal matrix.

Assuming half-filling and $U_\mathrm{el}\gg t_\mathrm{el}$, the onsite Coulomb repulsion penalizes double occupancy, and the low-energy subspace consists of states with exactly one electron per site.
In this low-energy subspace, charge degrees of freedom are frozen and only spin dynamics remain.
Using second-order perturbation theory,
virtual hopping processes generate a low-energy effective Hamiltonian, which is described by the $S=1/2$ antiferromagnetic Heisenberg model.
While this procedure is standard, the presence of the Peierls phase induced by spin–orbit coupling modifies the resulting Heisenberg Hamiltonian, which is ultimately given by
\begin{equation}
    H_S(E)=J\sum_i\frac{1}{2}(e^{-i\theta_\mathrm{AC}}S_i^-S_{i+1}^++e^{i\theta_\mathrm{AC}}S_i^+S_{i+1}^-)+S_i^zS_{i+1}^z,\label{eq:spin Hamiltonian with AC phase}
\end{equation}
where $J=4t_\mathrm{el}^2/U_\mathrm{el}>0$ and $S_i^\pm=S_i^x\pm iS_{i}^y$. 
The $S=1/2$ spin operators are defined in terms of the electron field operators as 
\begin{equation}
    \vb{S}_i=\frac{1}{2}\sum_{\sigma,\sigma'}c_{i\sigma}^\dagger\bm{\sigma}_{\sigma\sigma'}c_{i\sigma'}
\end{equation}
where $\bm{\sigma}=(\sigma^x,\sigma^y,\sigma^z)$ are Pauli matrices.
Explicitly, the spin components are given by
$S_i^+=c_{i\uparrow}^\dagger c_{i\downarrow},
~S_{i}^-=c_{i\downarrow}^\dagger c_{i\uparrow},
~S_i^z=(n_{i\uparrow}-n_{i\downarrow})/2$.

The AC phase in Eq.~\eqref{eq:spin Hamiltonian with AC phase} can be interpreted as originating from the Peierls phase accumulated by electrons during virtual hopping processes.
For example, 
the first term in Eq.~\eqref{eq:spin Hamiltonian with AC phase} corresponds to
an up-spin electron hopping from site $i$ to $i+1$, together with a down-spin electron hopping from site $i+1$ to $i$~[Fig.~\ref{figure: superexchange}(a)]. 
Each hopping process acquires a phase factor $e^{-i\theta_\mathrm{AC}/2}$, resulting in a total phase $e^{-i\theta_\mathrm{AC}}$.
The second term in Eq.~\eqref{eq:spin Hamiltonian with AC phase} can be analyzed in exactly the same manner [Fig.~\ref{figure: superexchange}(b)].

As a result, the Aharonov–Casher phase enters the Heisenberg Hamiltonian~[Eq.~\eqref{eq:spin Hamiltonian with AC phase}], thereby encoding the coupling between the spin and the electric field.
By taking a derivative with respect to the electric field, we obtain the general expression for the electric polarization
\begin{equation}
    \vb{P}_{i,i+1}=-\pdv{H_S}{\vb{E}},
\end{equation}
which remains valid in all order in electric field.

It is noteworthy that,
to first order in the electric field, 
the correction to the Hamiltonian~[Eq.~\eqref{eq:spin Hamiltonian with AC phase}] takes the form of the Dzyaloshinskii–Moriya (DM) interaction,
\begin{align}
    H_S(E)&\simeq H_S(0)+\frac{J}{2}\sum_i(-i\theta_\mathrm{AC}S_i^-S_{i+1}^++i\theta_\mathrm{AC}S_i^+S_{i+1}^-)\notag\\
    &=H_S(0)+\sum_i\vb{D}_\mathrm{DM}\cdot(\vb{S}_i\times\vb{S}_{i+1}),
\end{align}
where we define the DM vector as $\vb{D}_\mathrm{DM}=J\theta_\mathrm{AC}\vb{e}_z$.
This DM term can be recast in a form $-\sum_i \vb{P}_{i,i+1}\cdot\vb{E}$, giving the expression for the electric dipole to the zeroth order in the electric field~\cite{Katsura2005,Liu2011}
\begin{equation}
    \vb{P}_{i,i+1}\simeq g_\mathrm{AC}^{}\frac{J}{\hbar}(\vb{S}_i\times\vb{S}_{i+1})^z~\vb{e}_z\times\bm{a}_{i,i+1},
\end{equation}
where $\bm{a}_{i,i+1}=a\vb{e}_x$ is the lattice vector that connects site $i$ and $i+1$.
The magnon current operator can be formally defined in this system~\cite{Matsubara1956,Zapf2014}.
To zeroth order in the electric field, it is given by
$j_{i,i+1}\simeq-J/\hbar~(\vb{S}_i\times\vb{S}_{i+1})^z$.
Accordingly, the polarization $P=P_{i,i+1}/a^3$ is related to the magnon current density $j_\mathrm{m}=j_{i,i+1}/a^2$ via the general relation Eq.~\eqref{eq:P}.
Thus, even a static noncollinear spin texture corresponds to an equilibrium magnon current and consequently induces electric polarization.

\begin{figure}[t]
    \centering
    \includegraphics{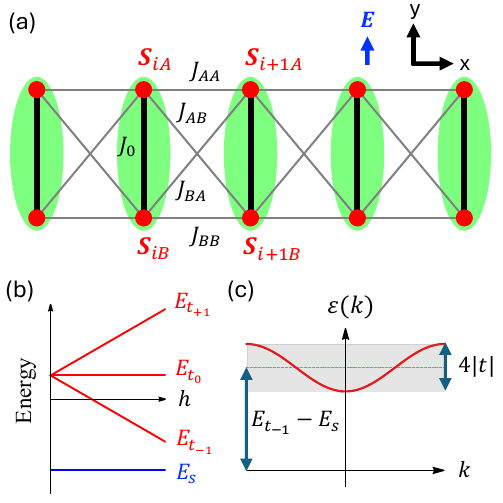}
    \caption{
    (a) Schematic illustration of a spin-1/2 antiferromagnetic dimer lattice. Each site hosts a localized spin $\vb{S}_{i\alpha}$ (red dots), and two spins within a dimer $i$ are labeled by sublattices $\alpha=A,B$.
    The intradimer antiferromagnetic exchange interaction $J_0$ is represented by thick black lines, while the weaker interdimer exchange couplings $J_{\alpha\beta}$ are shown as thin gray lines connecting neighboring dimers.
    (b) Magnetic-field dependence of the energy levels of the dimer eigenstates. The singlet state $E_s$ and triplet states $E_{t_m}$ ($m = 0, \pm1$) are split by the Zeeman effect. 
    As the magnetic field $h$ increases, $E_{t_{-1}}$ decreases and eventually approaches $E_s$.
   (c) Schematic magnon dispersion $\varepsilon(k)$. The originally flat triplet excitation acquires a finite bandwidth $4|t|$ due to the interdimer exchange interactions, which enable triplon hopping between dimers.
    }
    \label{figure: Dimer magnet}
\end{figure}
\subsection{Magnon Bose-Hubbard model}\label{subapp: spin model}
We consider a simple spin model~[Fig.~\ref{figure: Dimer magnet}(a)],
which yields the one-dimensional magnon Bose–Hubbard model~[Eq.~\eqref{eq: Bose-Hubbard model}] in the main text as an effective low-energy description.
Despite its simplicity, this spin model captures the essential ingredients for magnetic-field-induced magnon BEC~\cite{Nikuni2000,Ruegg2003,Giamarchi2008,Aczel2009,Zapf2014,Kimura2016}--
each dimer consists of two spins coupled by a strong antiferromagnetic exchange interaction,
while neighboring dimers are coupled via weaker interdimer interactions.

The Hamiltonian for the simple spin model is given by 
\begin{align}
    H_\mathrm{d}=H_\mathrm{intra}+H_\mathrm{inter}.
\end{align}
Here, the intradimer exchange term is defined as
\begin{align}
    H_\mathrm{intra}&=J_0\sum_{i} \vb{S}_{iA}\cdot\vb{S}_{iB}+h\sum_i(S^z_{iA}+S^z_{iB}),
\end{align}
where $J_0>0$ is the antiferromagnetic exchange coupling within a dimer, $h$ is the Zeeman magnetic field, and $\vb{S}_{i\alpha}$ is the spin-1/2 operator at site $\alpha=A,B$ of dimer $i$.
The interdimer exchange term describes couplings between spins belonging to different dimers:
\begin{align}
    H_\mathrm{inter}=\sum_i \sum_{\alpha,\beta}J_{\alpha\beta}\vb{S}_{i\alpha}\cdot\vb{S}_{i+1,\beta}
\end{align}
where $J_{\alpha\beta}>0$ represents antiferromagnetic exchange coupling between spin $\alpha$ in dimer $i$ and spin $\beta$ in dimer $i+1$.
\\
\indent
It is important to emphasize that the intradimer exchange is much stronger than the interdimer coupling,
\begin{equation}
    J_0\gg J_{\alpha\beta},
\end{equation}
which establishes a clear separation of energy scales.
As a consequence, as we will show in the remainder of this section, the intradimer interaction dominates the local physics, stabilizing a singlet ground state on each dimer, while the weaker interdimer coupling acts as a perturbation that enables triplet excitations to propagate through the lattice.
In the presence of a magnetic field, the Zeeman term splits the triplet manifold, allowing for a field-induced population of low-energy triplon excitations.

First of all,
we focus on a local unit (e.g., a dimer $i$) described by the Hamiltonian
\begin{align}
    H_\mathrm{intra}^{(i)}&=
    J_0\vb{S}_{iA}\cdot\vb{S}_{iB}+h(S^z_{iA}+S^z_{iB}),\notag\\
    &=\frac{J_0}{2}(S_{iA}^-S_{iB}^++S_{iA}^+S_{iB}^-)
    +J_0S_{iA}^zS_{iB}^z
    +h(S^z_{iA}+S^z_{iB})
\end{align}
The local energy spectrum consists of a singlet state and three triplet states, which are given by
\begin{align}
    \ket{s}&=\frac{\ket{\uparrow\downarrow}-\ket{\downarrow\uparrow}}{\sqrt{2}},~
    \ket{t_{-1}}=\ket{\downarrow\downarrow},\notag\\
    \ket{t_0}&=\frac{\ket{\uparrow\downarrow}+\ket{\downarrow\uparrow}}{\sqrt{2}},~
    \ket{t_{+1}}=\ket{\uparrow\uparrow}.
\end{align}
Here, $\ket{s}$ denotes the spin singlet, while $\ket{t_m}$ denote the triplet states with total spin projection $S_i^z\equiv S_{iA}^z+S_{iB}^z=m$.
Their energy eigenvalues are given by
\begin{align}
    E_s&=-\frac{3}{4}J_0,~E_{t_{-1}}=\frac{1}{4}J_0-h,\notag\\
    E_{t_{0}}&=\frac{1}{4}J_0,~E_{t_{+1}}=\frac{1}{4}J_0+h,
\end{align}
respectively.
In zero field, the singlet is the ground state, while the triplet states are degenerate excited states. 
Upon introducing a magnetic field ($h>0$), the triplet manifold is split by the Zeeman coupling, and the energy of the $S^z=-1$ triplet decreases linearly with increasing field [Fig.~\ref{figure: Dimer magnet}(b)].
In this regime, the low-energy Hilbert space is effectively spanned by $\ket{s}$ and $\ket{t_{-1}}$. 
It is then convenient to introduce a bosonic quasiparticle, referred to as a magnon (or triplon), by identifying $\ket{s}$ as the vacuum and $\ket{t_{-1}}$ as a single-particle excitation, giving
\begin{align}
    \ket{s}=b_i\ket{t_{-1}},~\ket{t_{-1}}=b_i^\dagger\ket{s}.
\end{align}
Here, $b_i^\dagger~(b_i)$ creates (annihilates) a magnon at site $i$, which is defined as
\begin{align}
    b_i^\dagger=\frac{1}{\sqrt{2}}(S_{iA}^--S_{iB}^-),~
    b_i=\frac{1}{\sqrt{2}}(S_{iA}^+-S_{iB}^+),\label{eq: definition of magnon}
\end{align}
respectively.
Within this reduced Hilbert space, the triplon obeys hardcore boson statistics, reflecting the local constraint that each site can host at most one excitation. 
This implies an infinite on-site repulsive interaction.

Next,
we focus on the interdimer coupling between the dimer $i$ and $i+1$, which is described by the Hamiltonian
\begin{align}
    H_\mathrm{inter}^{(i)}=&\sum_{\alpha,\beta}J_{\alpha\beta}\vb{S}_{i,\alpha}\cdot\vb{S}_{i+1,\beta}\notag\\
    =&\sum_{\alpha,\beta}\frac{J_{\alpha\beta}}{2}(S_{i,\alpha}^-S_{i+1,\beta}^+
    +S_{i,\alpha}^+S_{i+1,\beta}^-)
    +J_{\alpha\beta}S_{i,\alpha}^zS_{i+1,\beta}^z\label{eq: interdimer Hamiltonian i and i+1}
\end{align}
The first term in Eq.~\eqref{eq: interdimer Hamiltonian i and i+1} can be decomposed as
\begin{align}
    \sum_{\alpha,\beta}&\frac{J_{\alpha\beta}}{2}S_{i,\alpha}^-S_{i+1,\beta}^+\notag\\
    &=\frac{J_{AA}-J_{AB}-J_{BA}+J_{BB}}{8}(S_{iA}^--S_{iB}^-)(S_{i+1A}^+-S_{i+1B}^+)\notag\\
    &+\frac{J_{AA}+J_{AB}+J_{BA}+J_{BB}}{8}(S_{iA}^-+S_{iB}^-)(S_{i+1A}^++S_{i+1B}^+)\notag\\
    &+\frac{J_{AA}+J_{AB}-J_{BA}-J_{BB}}{8}(S_{iA}^--S_{iB}^-)(S_{i+1A}^++S_{i+1B}^+)\notag\\
    &+\frac{J_{AA}-J_{AB}+J_{BA}-J_{BB}}{8}(S_{iA}^-+S_{iB}^-)(S_{i+1A}^+-S_{i+1B}^+).
\end{align}
Only the first term contributes within the low-energy subspace spanned by $\ket{s}$ and $\ket{t_{-1}}$,
and this term corresponds to a hopping process of magnons.
The same analysis applies to the second term in Eq.~\eqref{eq: interdimer Hamiltonian i and i+1}, leading to
\begin{equation}
    H_\mathrm{inter}^{(i)}\rightarrow-t(b_i^\dagger b_{i+1}+b_i b_{i+1}^\dagger)\label{eq: magnon hopping term}
\end{equation}
where the hopping parameter is defined as
\begin{equation}
    -t\equiv\frac{J_{AA}-J_{AB}-J_{BA}+J_{BB}}{4},\label{eq: definition of -t}
\end{equation}
and the dispersion relation of the magnon is given by [Fig.~\ref{figure: Dimer magnet}(c)]
\begin{equation}
    \varepsilon(k)=(E_{t_{-1}}-E_s)-2t\cos(ka).
\end{equation}
It is important to note that, although we assume $t>0$ in the main text (for which the magnon band minimum is located at $k=0$),
all results, including the emergence of ferroelectricity, remain valid for $t<0$,
where the band minimum shifts to $k=\pi$.
\\
\indent
Let us consider the situation, where the system interacts with the electric field $\vb{E}=E\vb{e}_y$ through the spin-orbit coupling.
As discussed in Appendix.~\ref{subapp: AC phase},
each antiferromagnetic exchange coupling constituting the dimer lattice~[Fig.~\ref{figure: Dimer magnet}~(a)] acquires an AC phase, as given in Eq.~\eqref{eq:spin Hamiltonian with AC phase}.
However, since the electric field is assumed to be oriented along the $y$-direction, only the interdimer exchange couplings $J_{\alpha\beta}$ acquire the AC phase, while the intradimer coupling $J_0$ remains unaffected.
As a result, 
the nearest-neighbor magnon hopping parameter $-t$ acquires the AC phase through Eq.~\eqref{eq: definition of -t}, 
leading to the magnon Bose-Hubbard model with AC phase~[Eq.~\eqref{eq: Bose-Hubbard model}] in the main text. 

Then, we discuss how a magnetic field drives the system into a Bose–Einstein condensed phase of magnons.~\cite{Nikuni2000,Ruegg2003,Giamarchi2008,Aczel2009,Zapf2014,Kimura2016}
In the absence of a magnetic field, the ground state of the spin-dimer system is the singlet state, separated by an energy gap from the triplet excitations. 
The triplet (triplon) excitations form dispersive bands due to inter-dimer exchange interactions.
As the magnetic field increases, the minimum of the dispersion decreases. At a critical field 
\begin{equation}
    h_{c1}=J_0-2|t|,
\end{equation}
the gap closes.
This point marks the onset of magnon BEC,
where the lowest-energy triplon mode becomes macroscopically occupied, leading to a finite expectation value $\expval{b_i}\ne0$.
As follows from Eq.~\eqref{eq: definition of magnon}, it corresponds to the development of an in-plane staggered magnetization perpendicular to the applied magnetic field.
As the magnetic field is increased further, the triplon density grows until all spins become fully polarized. 
At a second critical field 
\begin{equation}
    h_{c2}=J_0+2|t|,
\end{equation}
the system reaches saturation, where  the top of the triplon band touches zero energy.
Thus, the field-induced magnon BEC exists only within the window
$h_{c1}<h<h_{c2}$.
In summary, below $h_{c1}$ the system is in a quantum disordered phase with a singlet ground state. 
Between $h_{c1}$ and $h_{c2}$, the system realizes a magnon condensate with long-range coherence.
Above $h_{c2}$, the system enters a fully polarized phase, where the bosonic description in terms of magnons is no longer appropriate.

Finally, let us clarify the correspondence between the magnon and spin representations. 
In the magnon BEC, the expectation value of the bosonic operator $\expval{b_i}$ becomes nonzero, signaling the spontaneous breaking of the U(1) symmetry.
In the spin language, using Eq.~\eqref{eq: definition of magnon}, this corresponds to the emergence of magnetic order in the plane perpendicular to the Zeeman field, i.e., an in-plane staggered magnetization,
which represents the spontaneous breaking of the SO(2) rotational symmetry.
\\\indent
We next consider a magnon current and identify its counterpart in the spin representation. 
The general expression for the magnon current is given in Eq.~\eqref{eq:current operator}
in terms of the bosonic operator $b_i$.
Focusing on the zeroth-order contribution in the electric field, it can be simply written as
\begin{align}
    j_{i,i+1}
    &\simeq\frac{it}{\hbar}(b_ib_{i+1}^\dagger-b_i^\dagger b_{i+1})\notag\\
    &=\frac{t}{\hbar}\Bigl[(\vb{S}_{iA}-\vb{S}_{iB})\times(\vb{S}_{i+1A}-\vb{S}_{i+1B})\Bigr]^z,\label{eq: zeroth order magnon current}
\end{align}
where we have used Eq.~\eqref{eq: definition of magnon} to express $j_{i,i+1}$ in terms of spin operators.
Therefore, a finite expectation value of the magnon current implies that the in-plane staggered magnetization develops a coherently twisted but static spin texture.

\section{Magnonic Bohr-van Leeuwen theorem}\label{sec: Bohr-van Leuwen}
The original Bohr-van Leeuwen theorem states that a classical electronic system in thermal equilibrium cannot sustain a net magnetization.
This highlights that magnetism cannot be explained by classical physics alone and must therefore arise from inherently quantum mechanical effects.

Here, we develop a magnonic analogue of the Bohr-van Leeuwen theorem: classical magnonic systems in thermal equilibrium are incapable of producing a net electric polarization.
The derivation is given as follows.
In classical mechanics, the motion of magnons in an electric field $\vb{E}(\vb{r})$ is governed by the Hamiltonian
\begin{equation}
    H(\vb{r},\bm{p})=\frac{(\bm{p}+\vb{A}_\mathrm{m}(\vb{r}))^2}{2m^*},
\end{equation}
that couples with the electric fields through the effective vector potential $\vb{A}_\mathrm{m}(\vb{r})=g^{}_\mathrm{AC}\vb{E}(\vb{r})\cross\vb{e}_z$.
Based on this Hamiltonian, we evaluate the electric dipole moment
$\vb{p}=-\partial H(\vb{r},\bm{p})/\partial \vb{E}$
within the framework of classical statistical mechanics.
Taking a thermal average in the phase space $(\vb{r},\bm{p})$, we get
\begin{align}
    \expval{\vb{p}}
    &\propto\int d\vb{r}~ d\bm{p}~g^{}_\mathrm{AC}\qty(\frac{\bm{p}+\vb{A}_\mathrm{m}(\vb{r})}{m^*}\times\vb{e}_z)\exp(-\frac{(\bm{p}+\vb{A}_\mathrm{m}(\vb{r}))^2}{2m^*k_BT})\notag\\
    &=\int d\vb{r}~ d(m^*\vb{v})~g^{}_\mathrm{AC}\qty(\vb{v}\times\vb{e}_z)\exp(-\frac{m^*\vb{v}^2}{2k_BT})\notag\\
    &=0,
\end{align}
where $1/(k_BT)$ is an inverse temperature. 
In the second equality, we performed a change of variables in the integration by introducing the mechanical momentum $m^*\vb{v}\equiv\bm{p}+\vb{A}_\mathrm{m}(\vb{r})$.
Through this change of variables, the explicit electric field dependence is removed, and the integral is shown to vanish identically.
This result demonstrates that the ferroelectric phase transition in magnon Bose-Einstein condensates presented in this work is attributed to intrinsically quantum mechanical origins.


\bibliography{reference}


\end{document}